\documentclass[twocolumn]{article}
\usepackage[a4paper, top=2cm, bottom=2cm, left=2cm, right=2cm]{geometry}
\usepackage{graphicx} 

\usepackage[labelfont=bf]{caption}

\usepackage[style=nature]{biblatex}

\usepackage{authblk}
\usepackage{amsmath} 
\usepackage{siunitx}
\usepackage{braket}
\usepackage{physics}
\usepackage{amssymb}
\usepackage{dsfont}
\usepackage[style=nature]{biblatex}

\usepackage{newfloat}
\DeclareFloatingEnvironment[name={Supplementary Figure}]{suppfigure}

\newcommand{\Bext}{\SI{335}{\milli\tesla}}
\newcommand{\GS}{\SI{1825.486503708015 \pm 0.3856392431363803}{\giga\hertz}}
\newcommand{\GSprime}{\SI{1111 \pm 86}{\giga\hertz}}

\newcommand{\wLe}{\SI{9.414}{\giga\hertz}}
\newcommand{\wLFree}{\SI{9.388}{\giga\hertz}}
\newcommand{\Finite}{\SI{0.86\pm0.01}{}}
\newcommand{\maxcontrast}{\SI{0.72\pm0.02}{}}

\newcommand{\TimeOne}{\SI{0.29589101140662677 \pm 0.08536659430645717}{\second}}
\newcommand{\cyc}{\SI{2019.2210334474744 \pm 381.1762320361209}{}}
\newcommand{\FigOneTimeTwoStar}{\SI{5.688203037479411 \pm 0.39876946062593266}{\micro\second}}
\newcommand{\FigOneAparaOneRam}{\SI{1193.563802026455 \pm 21.57925250607036}{}}
\newcommand{\FigOneAparaTwoRam}{\SI{418.34106355971715 \pm 21.57925250607036}{}}
\newcommand{\FigOneAparaThreeRam}{\SI{160.23151789415494 \pm 21.57925250607036}{}}

\newcommand{\FigCPTAparaOneCPT}{\SI{1143.5762367286829 \pm 30.754586308282378}{\kilo\hertz}}
\newcommand{\FigCPTAparaTwoCPT}{\SI{415.21143763602915 \pm 26.71381021963404}{\kilo\hertz}}
\newcommand{\FigCPTGammaCPT}{\SI{148.85492054619882 \pm 24.02657385251073}{\kilo\hertz}}

\newcommand{\FigOneTimeTwoHahn}{\SI{212.59716232213677 \pm 12.49692534892614}{\micro\second}}

\newcommand{\FigOneTimeTwoCPMG}{\SI{1.3116741089960194 \pm 0.1885018492583327}{\milli\second}}
\newcommand{\FigOneChi}{\SI{0.5133516863783796 \pm 0.026017217569571194}{}}
\newcommand{\FigDEERnuBath}{\SI{9.38885309702593 \pm 0.0037562414131802266}{\giga\hertz}}
\newcommand{\FigDEERnuSecondSiV}{\SI{9.43063987 \pm 0.000771273697}{\giga\hertz}}

\newcommand{\FigTwoDDTauOne}{\SI{8.395}{\micro\second}}
\newcommand{\FigTwoDDNOne}{\SI{48}{}}
\newcommand{\FigTwoDDTauTwo}{\SI{8.397}{\micro\second}}
\newcommand{\FigTwoDDNTwo}{\SI{92}{}}

\newcommand{\FigTwoomegaLnDD}{\SI{3.5825184138330926 \pm 0.000011300884786737325}{\mega\hertz}}
\newcommand{\FigTwoAperpOneDD}{\SI{233.19180977154946 \pm 0.8132586168936415}{}}
\newcommand{\FigTwoAperpTwoDD}{\SI{147.74748531557418 \pm 1.5225699183946046}{}}
\newcommand{\FigTwoAperpThreeDD}{\SI{75.51727008466486 \pm 3.5442021063133513}{}}
\newcommand{\FigTwoAperpFourDD}{\SI{46.815467374385136 \pm 4.430273142660529}{}}

\newcommand{\FigTwoDDNCoh}{\SI{274 \pm 17}{}}
\newcommand{\FigTwoTauTimeTwo}{\SI{8.43}{\micro\second}}
\newcommand{\FigTwoTimeTwo}{\SI{2.3108671159555234 \pm 0.14379712964219984}{\milli\second}}

\newcommand{\FigTwoTauInit}{\SI{8.3915}{\micro\second}}

\newcommand{\FigTwoNCXTwo}{\SI{24}{}}
\newcommand{\FigTwoTauCXTwo}{\SI{8.395}{\micro\second}}
\newcommand{\FigTwoAparaOneNucRamsey}{\SI{1110.9612409409253 \pm 48.15616441129946}{\kilo\hertz}}

\newcommand{\FigThreeTimeTwoLowPowerRabi}{\SI{51.28228927 \pm 10.11733808}{\micro\second}}
\newcommand{\FigThreeOmegaLowPowerRabi}{\SI{70.92898936 \pm 0.59207054}{\kilo\hertz}}
\newcommand{\FigThreeOmegaUltraLowPowerRabi}{\SI{5.492926850010497 \pm  0.2187926426197725}{\kilo\hertz}}
\newcommand{\FigThreeTimeTwoUltraLowPowerRabi}{\SI{161.02676146148752  \pm  29.523818746051766}{\micro\second}}

\newcommand{\FigThreeAparaOneODMR}{\SI{1181.920899126053 \pm 7.101240839587634}{}}
\newcommand{\FigThreeAparaTwoODMR}{\SI{411.535851146698 \pm 7.079668406612841}{}}
\newcommand{\FigThreeAparaThreeODMR}{\SI{144.61076615715027 \pm 7.0643156840511665}{}}
\newcommand{\FigThreeGammaODMR}{\SI{134.791080200405 \pm 2.238639740339079}{\kilo\hertz}}

\newcommand{\FigThreeSwapODMRRabi}{\SI{7.7}{\kilo\hertz}}
\newcommand{\FigThreeAparaFourODMR}{\SI{32.483925357818606 \pm 1.4704819455837308}{\kilo\hertz}}
\newcommand{\FigThreeGammaSwapODMR}{\SI{42.9968715042213 \pm 2.3490809337488113}{\kilo\hertz}}

\newcommand{\FigThreeGammaRamsey}{\SI{55.9596561667113 \pm 3.9230318888723184}{\kilo\hertz}}

\newcommand{\FigThreeBeatRabiOne}{\SI{19.5 \pm .25}{\kilo\hertz}}
\newcommand{\FigThreeBeatRabiTwo}{\SI{43.0 \pm 1.4}{\kilo\hertz}}

\newcommand{\FigThreeRabiDetuning}{\SI{38.64594986458907  \pm  0.8532257043777719}{\kilo\hertz}}

\newcommand{\FigFourSwapBWTwo}{\SI{500}{\kilo\hertz}}
\newcommand{\FigFourSwapBWThree}{\SI{150}{\kilo\hertz}}

\newcommand{\FigFourSwapFidTwo}{\SI{0.96}{}}
\newcommand{\FigFourSwapFidThree}{\SI{0.73}{}}
\newcommand{\FigFourSwapFidFour}{\SI{0.53}{}}

\newcommand{\FigFourSwapRabiOneUp}{\SI{5.068742442669473 \pm 0.017499667575871333}{\kilo\hertz}}
\newcommand{\FigFourSwapRabiOneDown}{\SI{3.5642322525600014 \pm 0.013311149237441797}{\kilo\hertz}}

\newcommand{\FigFourSwapRabiAparaOne}{\SI{1194}{}}
\newcommand{\FigFourSwapRabiAperpOne}{\SI{242}{}}
\newcommand{\FigFourSwapRabiAparaTwo}{\SI{420}{}}
\newcommand{\FigFourSwapRabiAperpTwo}{\SI{174}{}}
\newcommand{\FigFourSwapRabiAparaThree}{\SI{141}{}}
\newcommand{\FigFourSwapRabiAperpThree}{\SI{98}{}}

\newcommand{\FigFourTimeTwoStarOne}{\SI{3.719544601208023 \pm 0.2472965}{}}
\newcommand{\FigFourTimeTwoStarTwo}{\SI{5.390041687 \pm 0.3067553668}{}}
\newcommand{\FigFourTimeTwoStarThree}{\SI{4.873313966135035 \pm 0.5394928563412566}{}}
\newcommand{\FigFourTimeTwoStarFour}{\SI{8.1464950965 \pm 8.9049252599}{}}

\newcommand{\FigFourSEDORCouplingTwoOne}{\SI{5.119870272800416 \pm 0.35511451033282293}{\hertz}}
\newcommand{\FigFourSEDORCouplingTwoThree}{\SI{28.56382160516395 \pm 0.3518650684628978}{\hertz}}
\newcommand{\FigFourSEDORCouplingOneThree}{\SI{19.711365046571963 \pm 1.2222769381038774}{\hertz}}

\newcommand{\FigFourSEDORTimeTwoTwoOne}{\SI{0.137474595630 \pm 0.03184489772392481}{\second}}
\newcommand{\FigFourSEDORTimeTwoTwoThree}{\SI{0.18051720381295747 \pm 0.03167535039152897}{\second}}
\newcommand{\FigFourSEDORTimeTwoOneThree}{\SI{0.0897871224471652 \pm 0.018830922567845337}{\second}}

\newcommand{\FigFiveTimeSSR}{\SI{5}{\milli\second}}
\newcommand{\FigFiveTimeLaser}{\SI{60}{\micro\second}}

\newcommand{\FigFiveFidOne}{\SI{0.98}{}}
\newcommand{\FigFiveFidTwo}{\SI{0.98}{}}
\newcommand{\FigFiveFidThree}{\SI{0.96}{}}
\newcommand{\FigFiveRegContrast}{\SI{0.5925840610775928 \pm 0.034022445428576686}{}}
\newcommand{\FigFiveAvgNInit}{\SI{0.9286229559483964 \pm 0.017771894272139377}{}}

\newcommand{\FigFiveFinitn}{\SI{0.96}{}}

\newcommand{\FigSixBellFidZeroZero}{\SI{0.6888812931856684 \pm 0.011482081220303}{}}

\newcommand{\FigSixBellFidRegPost}{\SI{0.94}{}}

\begin{document}
\title{Bipartite entanglement in a nuclear spin register mediated by a quasi-free electron spin}
\author[1]{Marco Klotz$^*$}
\author[1]{Andreas Tangemann$^*$}
\author[1,2]{David Opferkuch}
\author[1,2]{Alexander Kubanek$^\dagger$}
\affil[1]{Institute for Quantum Optics, Ulm University, Albert-Einstein-Allee 11, 89081 Ulm, Germany}
\affil[2]{Center for Integrated Quantum Science and Technology (IQST), Ulm University, Albert-Einstein-Allee 11, Ulm 89081, Germany}
\setcounter{Maxaffil}{0}
\renewcommand\Affilfont{\itshape\small}
\date{7th August 2025}

\twocolumn[
\begin{@twocolumnfalse}
	\maketitle
	\begin{abstract}
		Quantum networks will rely on photons entangled to robust, local quantum registers for computation and error correction. 
        We demonstrate control of and entanglement in a fully connected three-qubit $^{13}\mathrm{C}$ nuclear spin register in diamond. 
        The register is coupled to a quasi-free electron spin-1/2 of a silicon-vacancy center (SiV).
        High strain decouples the SiVs electron spin from spin-orbit interaction reducing the susceptibility to phonons at liquid helium temperature.
        As a result, the electron spin lifetime of hundreds of milli seconds enables sensing of nuclear-nuclear couplings down to few hertz. 
        To detect and control the register we leverage continuous decoupling using shaped, low-power microwave and direct radio frequency driving. 
        Furthermore, we implement a nuclear spin conditional phase-gate on the electron spin to mediate bipartite entanglement.  
        This approach presents an alternative to dynamically decoupled nuclear spin entanglement, not limited by the electron spin's 1/2 nature, opening up new avenues to an optically-accessible, solid-state quantum register.
        \vspace{0.5cm} 
	\end{abstract}
\end{@twocolumnfalse}
]
\def\thefootnote{*}\footnotetext{These authors contributed equally to this work}
\def\thefootnote{$\dagger$}\footnotetext{Corresponding author: alexander.kubanek@uni-ulm.de}

\section*{Introduction}
\begin{figure}[h]
    \centering
	\includegraphics[scale=0.7]{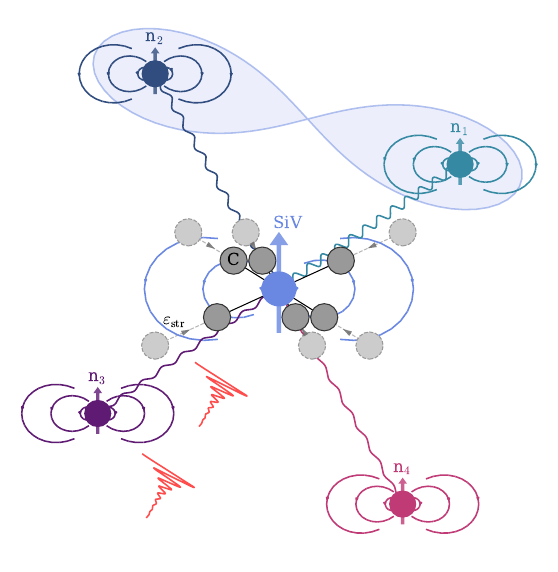}
	\caption{ 
    \textbf{Artistic sketch of a highly strained SiV with a quasi-free electron spin.}
    The quasi-free electron spin of a highly strained SiV (blue), with dashed and solid circles representing next-nearest carbon (C) atoms, is used as an optically addressable communication qubit.
    Hyperfine coupled and interconnected $^{13}$C nuclear spins n$_i$ are used as a quantum register with mutual entanglement (shaded blue area).     
    }	
	\label{fig:Intro}
\end{figure}
Solid-state quantum emitters demonstrated first implementations of quantum technologies \cite{awschalom2018quantum}, such as quantum sensors \cite{abobeih2019atomic, gottscholl2021spin} and quantum networks \cite{knaut2024entanglement, ruskuc2025multiplexed}. 
Color centers in diamond are of great interest for quantum networking applications due to exceptional optical and spin properties, demonstrating large spin registers with long memory times \cite{bradley2019ten, van2024mapping}, fault-tolerant information processing \cite{abobeih2022fault}, proof of principal error-corrected spin-photon entanglement \cite{chang2025hybrid}, distributed networking applications \cite{wei2025universal} and integration into scalable nanostructures \cite{rugar2021quantum}.\\
Group-IV color centers have superior spectral properties as compared to nitrogen-vacancy centers (NV).
However, local entanglement of highly coherent $^{13}$C nuclear spins, necessary for error detection and correction, has so far only been demonstrated with spin-1 NV \cite{chang2025hybrid}.\\
Very recently, the state-of-the-art approach of using dynamical decoupling (DD) of the defect's electron spin to control nearby nuclear spins \cite{nguyen2019quantum, Nguyen2019an} was paired with radio frequency control (DDRF) \cite{beukers2025control}, to demonstrate electron-nuclear entanglement and control of two $^{13}$C nuclear spins.  
In the case of group-IV centers, operation at \SI{1.7}{\kelvin} down to few mK is required to mitigate phonon-induced dephasing and magnetic field alignment with vector magnets is needed to prevent spin mixing \cite{karapatzakis2024microwave, rosenthal2023microwave, guo2023microwave, beukers2025control}. \\
Here, we demonstrate for the first time entanglement of two $^{13}$C nuclear spins coupled to a quasi-free electron spin of a negatively charged silicon-vacancy center (SiV).
We operate our highly-strained SiV at liquid helium temperature \cite{klotz2025ultra}.
We use efficient low-power microwave and radio frequency driving to directly detect and control up to four coupled nuclear spins and use high-fidelity state preparation with active feedback \cite{hesselmeier2024high}.
To establish nuclear-nuclear entanglement we utilize the geometric phase acquired during a $2\pi$-rotation of the  electron spin conditioned on three nuclear spins. \cite{waldherr2014quantum, xie2021beating}.
Our approach of detecting nuclear spins and generating entanglement relaxes constraints on the electron spin's hyperfine couplings and coherence time and can instead leverage the long $T^*_\mathrm{2,n}$ of the nuclear spins.

\section*{Results}
\subsubsection*{Spin characterization of a quasi-free electron}
\begin{figure}[h]
	\includegraphics[scale=1]{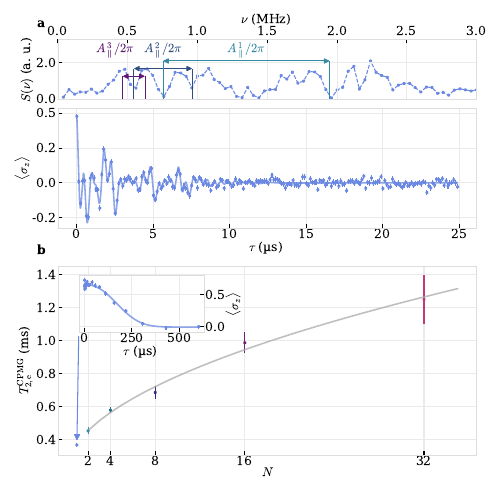}
	\caption{ \textbf{Electron spin coherence and nuclear spin coupling.} 
		\textbf{a} Detuned Ramsey interferometry with inter-pulse spacing $\tau$ on the electron spin (lower panel) with a corresponding Fourier spectrum $S(\nu)$ (upper panel). 
        Solid line is a fit to $\exp(-(\tau/T_\mathrm{2,e}^*)^\beta) \sum_{i=1}^{8} a_i\sin (\omega_i \tau + \phi^i)  + c$. \textbf{b} Scaling $\chi$ of the decoherence time $T_\mathrm{2,e}^\mathrm{CPMG}$ as a function of the number of decoupling pulses $N$, i.e. $T_\mathrm{2,e}^\mathrm{CPMG}\propto N^\chi$. 
        Inset shows a Hahn echo ($N =1$). 
        We extract the respective decay times from a fit of the form $a\exp (-(t/T_\mathrm{2,e})^\beta) + c$.    
    }    
	\label{fig:Figure1}
\end{figure}
\begin{figure*}[h]
	\includegraphics[scale=1]{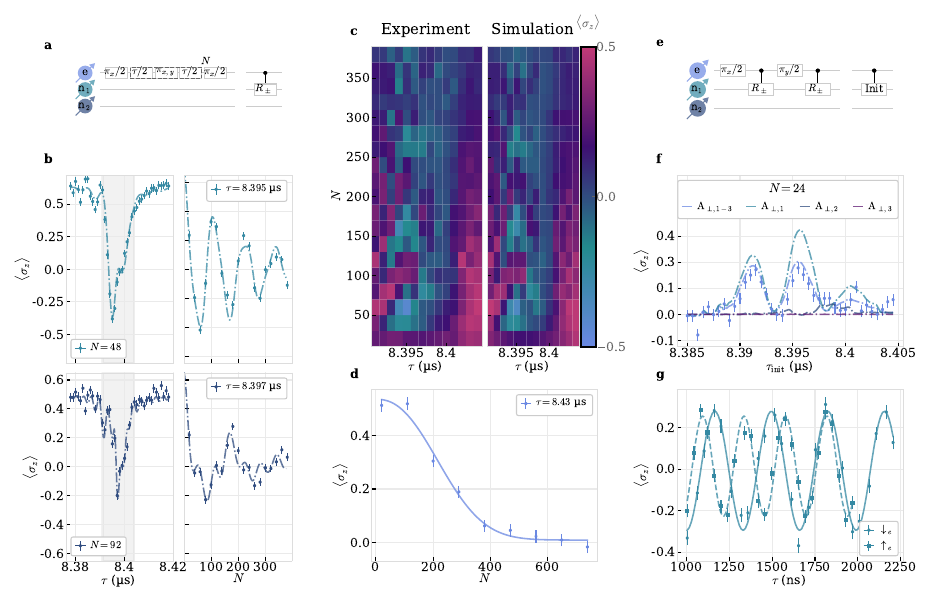}
	\caption{ \textbf{Dynamically decoupled nuclear spin detection and control.} 
		\textbf{a} 
        Pulse sequence for dynamically decoupled nuclear spin rotation conditional on the electron spin $R_\pm(N,\tau)$ \cite{taminiau2014universal}.         
        $\pi_\mathrm{x,y}$-elements are rotations around orthogonal axes, $\tau$-elements indicate free precession intervals and $N$ is the number of repetitions. 
        \textbf{b}
        An inter-pulse spacing $\tau$ sweep at $N = (\FigTwoDDNOne, \FigTwoDDNTwo)$, upper and lower left panel, shows dominant resonances of two nuclear spins, n$_1$ and n$_2$, where the electron spin's coherence is inverted at $\tau = (\FigTwoDDTauOne, \FigTwoDDTauTwo)$. 
        A subsequent sweep of $N$ at the resonant $\tau$ shows coherent oscillations, upper and lower right panel. 
        Dash-dotted lines are a numerical model.
        \textbf{c}
        A 2D ($\tau$, $N$) sweep around the two resonances from \textbf{b} showing the transition of strong coherent coupling to the weakly coupled bath. Left is the measured data from which we extract system parameters with a numerical model. The fitted model is shown on the right.
        \textbf{d}
        Off-resonant DD showing a loss of coherence after $N = \FigTwoDDNCoh$ ($T^\mathrm{XY}_\mathrm{2, e} = \FigTwoTimeTwo$), fitted with a stretched exponential decay, solid line.
        \textbf{e} 
        Nuclear spin initialization sequence. \cite{Nguyen2019an, maity2022mechanical} 
        \textbf{f} 
        Nuclear spin initialization with the pulse sequence depicted in \textbf{e}, where we construct a conditional effective rotation $R_\pm(\tau_\mathrm{init}, N)$, which incorporates a conditional x- and unconditional z-axis rotation of the nuclear spin \cite{taminiau2014universal, Nguyen2019an}.   
        Differently colored dash-dotted lines show a simulation for all nuclear spins (blue) and individual nuclear spins n$_1$ to n$_3$.
        \textbf{g} 
        Electron spin-dependent nuclear Ramsey interferometry confirming the hyperfine pairing $A_\parallel^1/2\pi = \FigTwoAparaOneNucRamsey$ and $A_\perp^1/2\pi=\FigTwoAperpOneDD\SI{}{\kilo\hertz}$ extracted from different measurements.
    }
	\label{fig:Figure2}
\end{figure*}
We are using a SiV hosted in a nanodiamond at liquid-helium temperatures, where strain $\epsilon_\mathrm{str}$ is largely exceeding the spin-orbit coupling $\lambda_\mathrm{SO} \approx \SI{50}{\giga\hertz}$ resulting in a ground-state splitting (GS) of $\Delta_\mathrm{GS} = \GS$.
Causes for high strain include differences in thermal expansion of the sapphire substrate and diamond \cite{berghaus2025cavity, guo2023microwave} as well as neighboring carbon vacancies \cite{xue2025all}, where the latter recently reported a SiV with a very similar ground-state splitting.
The properties of the sample have been described in detail in \cite{klotz2025ultra}.
Under such high GS, the SiV's electron spin becomes detached from its orbital degree of freedom, thus approaching a quasi-free electron spin \cite{Nguyen2019an, meesala2018strain}. 
This allows efficient electron spin driving and makes its relaxation time $T_\mathrm{1, e}$ robust against magnetic field misalignment \cite{meesala2018strain}.
Compared to recent results with a SiV with $\Delta_\mathrm{GS} = \GSprime$ \cite{klotz2025ultra}, we expect that phonon-induced decoherence and relaxation processes between orbital and spin states are further suppressed at our operating temperatures of $T \approx \SI{4}{\kelvin}$ and static magnetic field of $B_0 \approx \Bext$.
We polarize the electron spin optically, possible due to an anisotropic g-factor in the optical ground and excited state \cite{thiering2018ab}, thereby initializing the electron spin qubit with a fidelity of up to $F_\mathrm{e} \approx \Finite$. 
In each measurement we extract the fidelity from the amplitude $a$ and offset $c$ of a spin-pumping laser-pulse after inverting the electron spin population with an unconditional $\pi$-pulse, leading to a maximal achievable contrast of $\langle \sigma_z \rangle = \langle \uparrow \rangle - \langle \downarrow \rangle = 2 F_\mathrm{e}-1 = \maxcontrast$, where $F_\mathrm{e} = a/(a+c)$ \cite{klotz2025ultra}.\\
From power-dependent optical pump rates $\Gamma_\mathrm{p}$ we extract a cyclicity of $\eta = \cyc$.
We estimate a relaxation time of $T_\mathrm{1, e} = \TimeOne$, see \cite{SI}. \\
Compared to the less-strained SiV from \cite{klotz2025ultra}, this amounts to an increase by three orders of magnitude in $T_\mathrm{1, e}$, which is expected to also increase nuclear spins' coherence times, since uncontrolled electron spin flips induce decoherence due to different hyperfine precession frequencies.
We characterize the electron spin's coherence properties with coherent microwave (MW) control at a resonance frequency of $\wLe$, close to the resonance frequency of a free electron spin $\wLFree$. 
The Fourier transform of an off-resonant Ramsey experiment displays eight dominant frequency components, reminiscent of at least three coupled nearby nuclear spins n$_{1-3}$ with parallel hyperfine couplings $A^i_\parallel/2\pi = \left\{ \FigOneAparaOneRam, \FigOneAparaTwoRam, \FigOneAparaThreeRam \right\} \SI{}{\kilo\hertz}$ and an overall loss of coherence within $T^*_\mathrm{2,e} = \FigOneTimeTwoStar$, see Fig.\ref{fig:Figure1}a, setting the resolution of detecting nuclear spins.\\
We additionally perform a coherent population trapping (CPT) experiment and independently verify $A_\parallel^1 = \FigCPTAparaOneCPT$ and $A_\parallel ^2 = \FigCPTAparaTwoCPT$ with a resolution of $\Gamma_\mathrm{2,e}^\mathrm{CPT} = \FigCPTGammaCPT$, see \cite{SI}.\\
Using a single refocusing MW-pulse, i.e. a Hahn echo, the spin's coherence time can be extended to $T_\mathrm{2,e}^\mathrm{Hahn} = \FigOneTimeTwoHahn$, see inset of Fig.\ref{fig:Figure1}b. \\
Increasing the number of decoupling pulses to $N = 32$  in a CPMG type of measurement, the coherence time can be further extended to $T_\mathrm{2,e}^\mathrm{CPMG} = \FigOneTimeTwoCPMG$. 
The resulting scaling, $T^\mathrm{CPMG}_\mathrm{2, e} \propto  N^\chi $, is well described with a scaling factor $\chi = \FigOneChi$, shown in Fig.\ref{fig:Figure1}b. 
We attribute the dominant noise source to a bath of free electron spins, which we further investigate by performing double electron-electron resonance (DEER), see \cite{SI}.
We find a spectrally broad resonance at $\FigDEERnuBath$ which agrees well with the Larmor frequency of a free electron spin. 
Additionally, we find a more prominent and sharp resonance in the DEER spectrum at $\FigDEERnuSecondSiV$ which requires further investigation.
\subsubsection*{Dynamically decoupled nuclear spin detection and control}
In order to leverage highly coherent $^{13}$C nuclear spins from the surrounding lattice as a quantum register, individual detection and coherent control is required.
One approach relies on dynamically decoupling (DD) the electron spin from the surrounding nuclear spin bath except for specific target nuclear spins.
Under the right conditions, i.e. the right inter-pulse spacing $\tau$ and number of $\pi$-pulses $N$ in the pulse sequence depicted in Fig.\ref{fig:Figure2}a, entanglement of that nuclear spin with the electron spin leads to a loss of coherence \cite{Nguyen2019an,taminiau2014universal}. 
Due to the spin- 1/2 nature of the SiV's electron spin, this approach is only sensitive to second order in perpendicular hyperfine coupling, $A_\perp$, and nuclear Larmor frequency, $\omega_\mathrm{L,n}$, i.e. $(A_\perp/\omega_\mathrm{L,n})^2$ \cite{zahedian2024blueprint, takou2023precise}, therefore necessitating large $\tau$ to separate resonances of different nuclear spins, requiring long coherence time. \\
In Fig.\ref{fig:Figure2}b we sweep $\tau$ (left) and $N$ (right) of an XY-($\tau, N$) DD sequence around an exemplary resonance indicating entanglement with various nuclear spins. 
Clearly visible are the oscillations of the electron spin's coherence with $N$ at different $\tau = \FigTwoDDTauOne$ and $\FigTwoDDTauTwo$ for the most prominent resonances.\\
We extend the measurement to a full 2D sweep of $\tau$ and $N$ around the resonance and use a numerical model including the central electron spin, three target and one parasitic nuclear spin to fit the 2D dataset with good agreement to the experimental data, see Fig.\ref{fig:Figure2}c.
This allows us to extract the bare Larmor frequency $\omega_\mathrm{L,n}/2\pi = \FigTwoomegaLnDD$ as well as the perpendicular hyperfine coupling components of the strongest and parasitic spins $A^i_\perp/2\pi = \{ \FigTwoAperpOneDD, \FigTwoAperpTwoDD, \FigTwoAperpThreeDD, \FigTwoAperpFourDD\}\SI{}{\kilo\hertz}$. 
See \cite{SI} for further explanation of the used model and parameters. \\
In order to probe the electron spin's coherence under XY-$N$ DD we choose $\tau = \FigTwoTauTimeTwo$, off-resonant with a multiple of the nuclear spins precession period, sweep $N$ and observe an exponential decay in coherence within $T^\mathrm{XY}_\mathrm{2, e} = \FigTwoTimeTwo$, shown in Fig.\ref{fig:Figure2}d. \\
We verify exemplarily that the nuclear spin with $A_\perp ^1$ belongs to $A_\parallel^1$ by using a tailored DD initialization sequence from \cite{Nguyen2019an, maity2022mechanical, klotz2025ultra}, requiring electron spin conditioned $\pi/2$-rotations of the target nuclear spin as well as free precession intervals, depicted in Fig.\ref{fig:Figure2}e.
We implement the corresponding sequence with $N = \FigTwoNCXTwo$ and sweep $\tau_\mathrm{init}$ around the resonance, where individual initialization is achieved at $\tau_\mathrm{init,1} = \FigTwoTauInit$, see Fig.\ref{fig:Figure2}f.
Using the initialization sequence together with a conditional $\pi/2$-rotation, implemented with the pair ($\tau_\mathrm{CX} = \FigTwoTauCXTwo$, $N_\mathrm{CX} = \FigTwoNCXTwo$), we probe electron-spin dependent nuclear Ramsey interferences, depicted in Fig.\ref{fig:Figure2}g, which reveal two distinct precession frequencies split by $\FigTwoAparaOneNucRamsey$, in good agreement with $A_\parallel^1$ extracted from electron Ramsey measurements.\\
The procedure could be extended to the other two nuclear spins in a future study.
Addressing even more nuclear spins using DD sequences on the electron spin would require either a more favorable ratio of $A^i _\perp / \omega_\mathrm{L,n}$ or longer inter-pulse spacings to separate more weakly coupled nuclear spins from the bath. 
However, because of the high strain we rely on a strong magnetic field and therefore large $\omega_\mathrm{L,n}$ to still retain a high electron spin initialization fidelity. 
For this reason, we switch to a more direct approach to nuclear spin detection and control.
\subsubsection*{Direct nuclear spin spectroscopy}
\begin{figure}[h]
	\centering
	\includegraphics[scale=1]{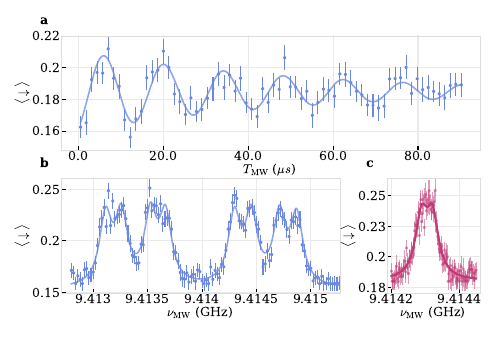}
	\caption{ \textbf{Low-power microwave sensing} 
		\textbf{a} 
        Continuously decoupled low-power Rabi driving showing persistent oscillations at a frequency of $\Omega_\mathrm{R,e}/2\pi=\FigThreeOmegaLowPowerRabi$ beyond the dephasing time $T_\mathrm{2,e}^* = \FigOneTimeTwoStar$. 
        Solid line is a fit to $a\sin (\Omega_\mathrm{R,e}T_\mathrm{MW}+\phi)\exp(-T_\mathrm{MW}/T_\mathrm{2,e}^\mathrm{Rabi}) + c$
        \textbf{b} 
        Frequency scan of a $\pi$-pulse with $\Omega_\mathrm{R,e}$ from \textbf{a} reveals eight resonances which are fit with $a\; \sum_{i=1}^8\mathcal{L}(\nu_\mathrm{MW}, \nu_0^i, \Gamma) + c$ (solid line), where $\mathcal{L}(\nu_\mathrm{MW}, \nu_0^i, \Gamma)$ is a Lorentzian.
        \textbf{c} 
        Polarizing the three nuclear spins from \textbf{b} before scanning the frequency at $\Omega_\mathrm{R,e} = \FigThreeSwapODMRRabi$ increases sensitivity below the unpolarized inhomogeneous linewidth and shows a fourth nuclear spin with $A_\parallel^4 = \FigThreeAparaFourODMR$, extracted from a double-Lorentzian (solid line).
    }
	\label{fig:Figure3}
\end{figure}
\begin{figure*}[h]
	\centering
	\includegraphics[scale=1]{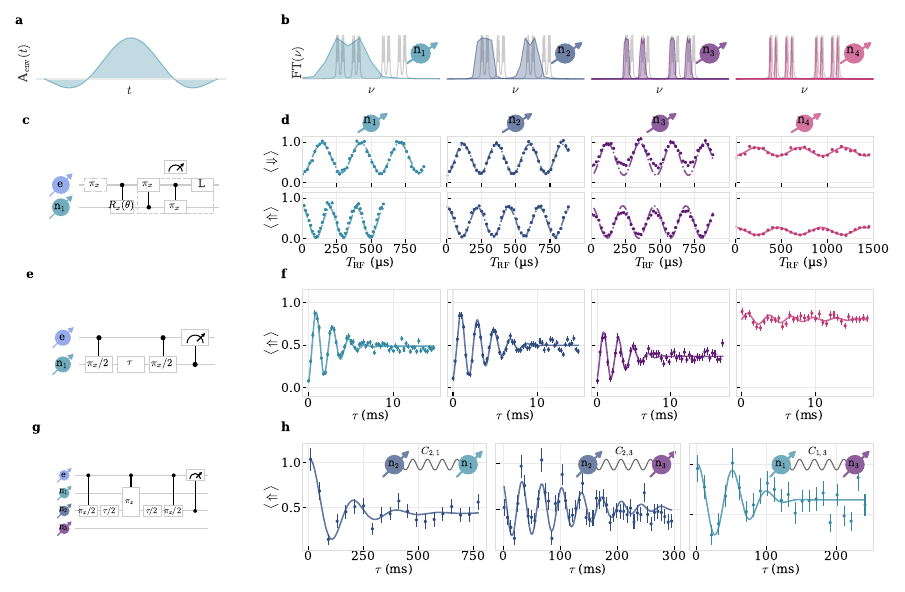}
	\caption{ \textbf{Radio frequency control and characterization of four nuclear spins.} 
		\textbf{a} Temporal and \textbf{b} spectral shape of the truncated sinc-pulses to realize the four different $\mathrm{C}_{\mathrm{n}_i}\mathrm{NOT}_\mathrm{e}$, where n$_i=\ket{\Downarrow_i}$.
        \textbf{c} Pulse sequence to rotate and swap a desired nuclear spin with the electron spin followed by a readout laser-pulse.
        \textbf{d} Electron spin-dependent Rabi oscillations of the populations $\ket{\Uparrow}$/$\ket{\Downarrow}$, read out with the respective $\mathrm{C}_{\Uparrow_i/\Downarrow_i}\mathrm{NOT}_\mathrm{e}$. 
        Upper/lower row indicate measurements where the electron is in $\ket{\uparrow}$/$\ket{\downarrow}$, necessary to realize both $\mathrm{C}_{\uparrow/\downarrow}\mathrm{NOT}_{\mathrm{n}_i}$ on all four nuclear spins. 
        Dash-dotted lines are obtained from simulations of our numerical model, see Methods. 
        Solid line is a fit to $a_i\sin (\Omega_\mathrm{R,n} \tau + \phi) \mathrm{exp}(-(t/T_\mathrm{2,Rabi,n})) + c$
        \textbf{e} Pulse sequence and \textbf{f} experimental data for nuclear Ramsey interferometry on all four nuclear spins with fits to $a \sin (\Delta \tau + \phi)\mathrm{exp}(-(\tau/T_\mathrm{2,n}^*)^\beta) + c$ (solid lines).
        \textbf{g} Pulse sequence for SEDOR experiment involving n$_2$ as sensor and n$_1$ as target spin.
        \textbf{h} Measurement of the mutual coupling $C_\mathrm{s,t}$ between the three strongest-coupled nuclear spins, with the respective sensor ($s$) and target ($t$) spin.
        Solid lines are fits to $a\sin (C_\mathrm{s,t} \tau + \phi)\mathrm{exp}(-(\tau/T_\mathrm{2,s}^\mathrm{SEDOR})^\beta) + c$.
    }
	\label{fig:Figure4}
\end{figure*}
The strain-induced strong decoupling of orbital and spin degree of freedom protects the electron spin's transition frequency from shifts due to strain fluctuations which can be present in nanohosts \cite{meesala2018strain, Nguyen2019an}.
In addition, the use of cryogenically cooled permanent magnets further reduces external magnetic field fluctuations, hence allowing us to drive Rabi oscillations with low-power MW-pulses while continuously decoupling. 
For example, Fig.\ref{fig:Figure3}a shows multiple Rabi oscillations with $\Omega_\mathrm{R,e}/2\pi = \FigThreeOmegaLowPowerRabi$ and $T_\mathrm{2,e}^\mathrm{Rabi} = \FigThreeTimeTwoLowPowerRabi$, an order of magnitude longer than $T^*_\mathrm{2,e}$. 
We can measure Rabi oscillations down to $\FigThreeOmegaUltraLowPowerRabi$ with $T_\mathrm{2,e}^\mathrm{Rabi} = \FigThreeTimeTwoUltraLowPowerRabi$, see \cite{SI}.\\
We scan for various nuclear spin-dependent resonances sweeping the frequency of a $\pi-$pulse with the previously determined Rabi frequency of $\Omega_\mathrm{R,e}/2\pi = \FigThreeOmegaLowPowerRabi$.
Fig.\ref{fig:Figure3}b shows a spectrum of eight distinct resonances of the previously found nuclear spins, also visible in Fig.\ref{fig:Figure1}a, from which we extract $A^i_{\parallel}/2\pi = \left\{ \FigThreeAparaOneODMR, \FigThreeAparaTwoODMR, \FigThreeAparaThreeODMR \right\}$ with a full-width-half-maximum (FWHM) of $\Gamma = \FigThreeGammaODMR > \Omega_\mathrm{R,e}/2\pi$, hinting towards the presence of a fourth nuclear spin which splits the eight resonances further but is undetectable.\\
Interestingly, if we polarize the three strongest nuclear spins, see next section, and perform the same measurement with a lower Rabi frequency of $\Omega_\mathrm{R,e}/2\pi = \FigThreeSwapODMRRabi$ the resolution increases and we can tentatively assign a fourth splitting with $A_\parallel^4 = \FigThreeAparaFourODMR$ with a width of $\FigThreeGammaSwapODMR$ which is below the inhomogeneous linewidth $\Gamma_\mathrm{2,e}^*/2\pi = C(\beta)\cdot 1/(\pi T_\mathrm{2,e}^*) = C(\beta) \cdot \FigThreeGammaRamsey$, where $C(\beta)=1(  2\sqrt{\ln 2})$ for a stretch factor $\beta=1(2)$ in the Ramsey measurement \cite{SI}. \\
A reduction in dephasing rate upon polarisation of the nuclear spin bath has also been observed for NV- centers in diamond \cite{london2013nv} and quantum dots \cite{jackson2022optimal}. 
Moreover, we verify the presence of a fourth nuclear spin by resolving a beat tone in a low-power Rabi measurement with resonant and effective Rabi frequency $\Omega_\mathrm{R, e}/2\pi =\FigThreeBeatRabiOne$ and $\Omega^\mathrm{eff}_\mathrm{R, e}/2\pi = \FigThreeBeatRabiTwo$, reminiscent of a detuning of $\Delta/2\pi = \FigThreeRabiDetuning$, close to the $A_\parallel^4/2\pi$ \cite{SI}.
\subsubsection*{Radio frequency nuclear spin control}
\begin{figure}[h]
	\centering
	\includegraphics[scale=1]{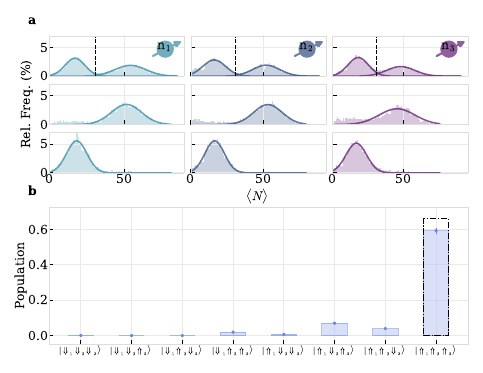}
	\caption{ \textbf{Single shot nuclear spin readout} 
		\textbf{a} Photon counting statistics of repetitively applying a $\mathrm{C}_{\Uparrow_i/\Downarrow_i}\mathrm{NOT}_\mathrm{e}$ with a short laser for a total laser on time $T_\mathrm{SSR} = \FigFiveTimeSSR$. Top row shows bare histogram without feedback. Center and bottom row show distributions with feedback and active control of an RF $\pi$-pulse after the SSR window to prepare a desired nuclear spin state, where each column indicates n$_1$ to n$_3$ from left to right. 
        \textbf{b}      
        Deterministic preparation of register state, $\ket{\mathrm{\Uparrow_{_1}\Uparrow_{_2}\Uparrow_{_3}}}$ in this example, by applying a SSR to each n$_i$ with active feedback. 
        Readout on each resonance is done with a $\FigFourSwapBWThree$ sinc-pulse. 
        Dash dotted bar indicates the expected contrast with nuclear spin initializations as determined in (a). 
    }
	\label{fig:Figure5}
\end{figure}

Using the extracted $\omega_\mathrm{L,n}$ and $\{ A_\parallel^i, A_\perp^i \}$, we can search more directly for the coupled spins' resonances and coherently control them.
We realize a nuclear-spin controlled electron flip $\mathrm{C}_{\Uparrow_i/\Downarrow_i}\mathrm{NOT}_\mathrm{e}$ for all four nuclear spins by consecutively driving the respective combinations of electron spin resonances with sinc-shaped MW-pulses to increase spectral selectivity. 
We truncate the sinc-pulses at the second zero crossing to reduce the temporal extend of the pulses.
The bandwidths of the $\mathrm{C}_{\Uparrow_i/\Downarrow_i}\mathrm{NOT}_\mathrm{e}$ are chosen to be close to each $A_\parallel^i/2\pi$, respectively, see Fig.\ref{fig:Figure4}a, b and \cite{SI}.
To realize electron-spin controlled nuclear flips $\mathrm{C}_{\uparrow/\downarrow}\mathrm{NOT}_{\mathrm{n}_i}$ we use rectangular radio frequency (RF) pulses. \\
We initialize and read out each nuclear spin n$_i$ by swapping the electron and nuclear spins populations with a $\mathrm{C}_{\Uparrow_i}\mathrm{NOT}_\mathrm{e}$ and a $\mathrm{C}_{\uparrow}\mathrm{NOT}_{\mathrm{n}_i}$ before reading out the electron spin, see Fig.\ref{fig:Figure4}c. 
By inverting the electron spin with an optional $\pi$-pulse and changing the respective RF frequency, we can drive coherent nuclear Rabi oscillations with Rabi frequencies $\Omega_{\mathrm{R, n}_i}/2\pi$ from $\FigFourSwapRabiOneDown$ to $\FigFourSwapRabiOneUp$ on all four nuclear spins conditional on the electron spin $\ket{\uparrow}/\ket{\downarrow}$, circuit diagram of the sequence and measurement data are shown in Fig.\ref{fig:Figure4}c and d. 
The offset and decreasing contrast with decreasing $A_\parallel$ is attributed to limited $\mathrm{C}_{\Uparrow_i/\Downarrow_i}\mathrm{NOT}_\mathrm{e}$ fidelities ranging from $\FigFourSwapFidTwo$ for n$_2$ to $\FigFourSwapFidFour$ for n$_4$, most likely arising from either pulse errors due to a detuning or onset of decoherence since the pulses get longer with decreasing bandwidth. 
See \cite{SI} for measurement data of the sinc-pulses.
These interpretations can be further verified by considering simulations done with the previously established numerical model consisting of one electron and four nuclear spins, see dash-dotted lines in Fig.\ref{fig:Figure4}d.  
Additionally, we deliberately used a lower $\Omega_\mathrm{R,n}$ for n$_4$, since we observe a beat tone when driving with $\Omega_\mathrm{R, n_4}/2\pi \approx \SI{4}{\kilo\hertz}$ potentially from unwanted polarization and driving of a fifth nuclear spin.
Ramsey measurements with detuning $\Delta \approx\SI{500}{\hertz}$ on each n$_i$ yield typical dephasing times of $T_{\mathrm{2,n}_i}^* = \left\{\FigFourTimeTwoStarOne, \FigFourTimeTwoStarTwo, \FigFourTimeTwoStarThree, \FigFourTimeTwoStarFour \right\}\SI{}{\milli\second}$ \cite{beukers2025control, van2024mapping}, see Fig.\ref{fig:Figure4} e,f. 
We further investigate the connectivity of our nuclear spin register using spin-echo double resonance (SEDOR) measurements on the three strongest coupled nuclear spins \cite{van2024mapping}. 
We perform a Hahn echo on the sensor nuclear spin ($s$) to increase the coherence time and therefore sensitivity.
Simultaneous to the refocusing $\pi$-pulse on the sensor, we apply a $\pi$-pulse on the target nuclear spin ($t$). 
As a result, we can accumulate the phase from a mutual coupling $C_\mathrm{s,t}$ between the two spins, whereas every other nuclear spins' couplings are rephased. 
Fig.\ref{fig:Figure4}g shows exemplarily the sequence for $\mathrm{n_s} = \mathrm{n_2}$ and $\mathrm{n_t} = \mathrm{n_1}$.
The measured oscillations in the nuclear spin coherence, see Fig.\ref{fig:Figure4}h are a result of mutual coupling $C_\mathrm{s,t}$ between the spins. 
The coupling components and coherence times are listed in Table \ref{tab:SEDOR}. 
\begin{table}
    \centering
    \setlength{\tabcolsep}{15pt} 
    \renewcommand{\arraystretch}{1.3} 
    \begin{tabular}{c||c||c}
         s,t & $C_\mathrm{s,t}/2\pi$ & $T^\mathrm{SEDOR}_\mathrm{2,n}$ \\ \hline\hline
        \multicolumn{1}{r||}{2,1} & \multicolumn{1}{r||}{\FigFourSEDORCouplingTwoOne} & \multicolumn{1}{r}{\FigFourSEDORTimeTwoTwoOne} \\
        \multicolumn{1}{r||}{2,3} & \multicolumn{1}{r||}{\FigFourSEDORCouplingTwoThree} & \multicolumn{1}{r}{\FigFourSEDORTimeTwoTwoThree} \\
        \multicolumn{1}{r||}{1,3} & \multicolumn{1}{r||}{\FigFourSEDORCouplingOneThree} & \multicolumn{1}{r}{\FigFourSEDORTimeTwoOneThree}\\
    \end{tabular}
    \caption{Register connectivity measured with SEDOR. s,t : sensor/target nuclear spin. $C_\mathrm{s,t}$: coupling strength between s and t. $T^\mathrm{SEDOR}_\mathrm{2,n}$: Coherence time extracted from SEDOR measurement.}
    \label{tab:SEDOR}
\end{table}
Having determined all RF resonance frequencies $\omega_{\uparrow/\downarrow, \mathrm{n}_i}/2\pi = \sqrt{(\omega_\mathrm{L,n} \pm A_\parallel^i/2)^2 + (\pm A_\perp^i/2)^2}/2\pi$, we can independently extract the hyperfine parameters of the three strongest nuclear spins $A_\parallel^i/2\pi = \left\{ \FigFourSwapRabiAparaOne, \FigFourSwapRabiAparaTwo, \FigFourSwapRabiAparaThree\right\}\SI{}{\kilo\hertz}$ and $A_\perp^i/2\pi = \left\{ \FigFourSwapRabiAperpOne, \FigFourSwapRabiAperpTwo, \FigFourSwapRabiAperpThree\right\}\SI{}{\kilo\hertz}$ which is more accurate compared to the fit of the 2D XY-$N$ measurements, since we are here only limited by the $T_\mathrm{2,n}^*$ of the nuclear spins.\\
In order to improve and overcome the electron-spin limited initialization fidelities of each n$_i$ realized by the previous swap sequence, we implement initialization by measurement. 
We can readout each nuclear spin's state by applying the corresponding aforementioned $\mathrm{C}_{\Uparrow_i/\Downarrow_i}\mathrm{NOT}_\mathrm{e}$ together with a $\FigFiveTimeLaser$ readout laser-pulse and repeat it for a total laser-on time of $T_\mathrm{SSR} = \FigFiveTimeSSR$.
To prevent nuclear spin polarization \cite{hesselmeier2024high, klotz2025ultra} we apply the sequence alternatingly on both nuclear spin states by switching the $\mathrm{C}_{\Uparrow_i/\Downarrow_i}\mathrm{NOT}_\mathrm{e}$ frequencies from $\ket{\Downarrow_i}$ to $\ket{\Uparrow_i}$.
We then count the number of collected photons within the single-shot readout (SSR) window and make a histogram of the photon counts, see first row of Fig.\ref{fig:Figure5}a for nuclear spins $\mathrm{n_1}-\mathrm{n_3}$.
In this way we can discriminate dark and bright states by choosing a threshold, indicated in Fig.\ref{fig:Figure5}a by a dashed line, which minimizes the mutual overlap of two Gaussian distributions.
This leads to fidelities $F_{\mathrm{n}_i} \approx \{\FigFiveFidOne, \FigFiveFidTwo, \FigFiveFidThree\}$ for n$_{1-3}$ \cite{klotz2025ultra}. \\
Moreover, we can deterministically prepare arbitrary nuclear register states with active-feedback by controlling the application of an RF $\pi$-pulse on n$_i$ conditioned on the number of detected photons in a SSR window. We count the photons within a SSR windows with a FPGA counter which conditionally on the counted photon number blocks the application of a subsequent RF $\pi$-pulse.
In the second and third row of Fig.\ref{fig:Figure5}a the measured photon statistics for the two different $\mathrm{C}_{\Uparrow_i/\Downarrow_i}\mathrm{NOT}_\mathrm{e}$ readout frequencies are shown for each nuclear spin, where active feedback prepares $\mathrm{n_1}$-$\mathrm{n_3}$ in a specific state.
As an example, we prepare the register state $\ket{\mathrm{\Uparrow_{_1}\Uparrow_{_2}\Uparrow_{_3}}}$ and apply  multi-conditioned $\FigFourSwapBWThree$ $\pi$-pulses on the eight electron spin transitions, thereby reading out the register's state. 
For the $\ket{\mathrm{\Uparrow_{_1}\Uparrow_{_2}\Uparrow_{_3}}}$ resonance we measure a population of $\FigFiveRegContrast$, see Fig.\ref{fig:Figure5}b, which corresponds to an average nuclear spin initialization of $\FigFiveAvgNInit$ of each nuclear spin, corrected for the readout infidelity of the $\SI{150}{\kilo\hertz}$ pulse. 
We attribute the discrepancy to the measured individual fidelities $F_{\mathrm{n}_i}$ to instabilities of the experimental setup which lead to drifts in photon-collection efficiency and therefore misclassifications after the SSR windows.
\subsubsection*{Phase-gate and nuclear-nuclear spin entanglement}
\begin{figure*}[h]
	\centering
	\includegraphics[scale=1]{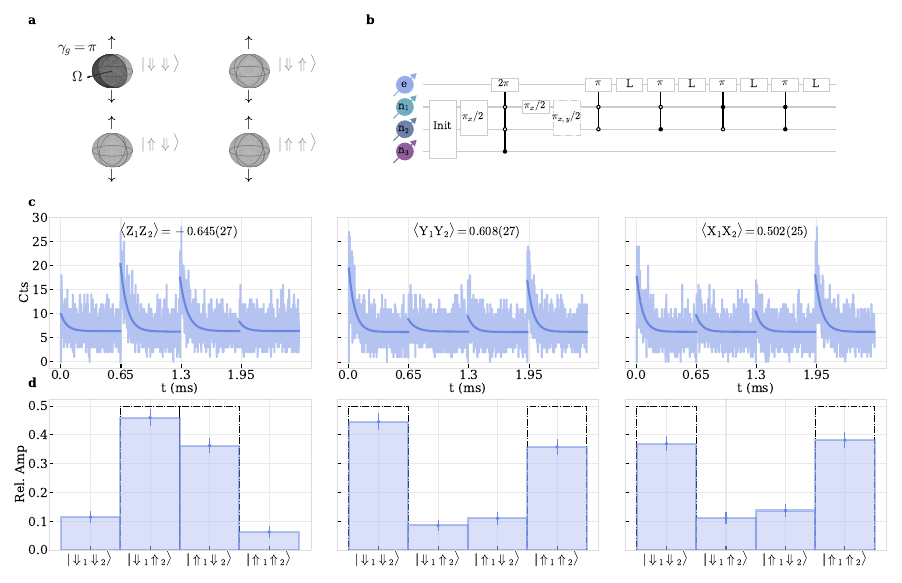}
	\caption{ \textbf{Nuclear-nuclear entanglement generation.} 
		\textbf{a} Geometric $\pi$-phase $\gamma_\mathrm{g}=\pi$ acquired during a full $2\pi$ evolution of the electron spin which encloses solid angle $\Omega = 2\pi$ and is conditioned on $\ket{\Downarrow_{_1} \Downarrow_{_2}}$.
        \textbf{b}
        Circuit diagram for generating a Bell-state between n$_1$ and n$_2$. 
        Init blocks consist of three SSR windows followed by an additional post-selection, see text. 
        $L$ is a read-out laser-pulse. 
        Full/hollow circles are conditional gates on $\ket{\Uparrow}/\ket{\Downarrow}$. 
        Dashed line indicate additional single-qubit gates to rotate the readout basis.
        \textbf{c}
        Photon histogram during read-out lasers of $\ket{\mathrm{\Downarrow_{n_1} \Downarrow_{n_2}}}$, $\ket{\mathrm{\Downarrow_{n_1} \Uparrow_{n_2}}}$, $\ket{\mathrm{\Uparrow_{n_1} \Downarrow_{n_2}}}$ and $\ket{\mathrm{\Uparrow_{n_1} \Uparrow_{n_2}}}$ resonances. 
        The different columns show the signal after application of additional basis rotations, see \textbf{b}.
        \textbf{d}
        Relative readout amplitudes of \textbf{c}, $a_i/\sum_i a_i$, reflecting the populations of respective nuclear spin states. 
        Black dash-dotted line show the populations and coherences of an ideal Bell-state $(\ket{\mathrm{\Downarrow_{n_1} \Uparrow_{n_2}}} + \ket{\mathrm{\Uparrow_{n_1} \Downarrow_{n_2}}})/\sqrt{2}$.
    }            
	\label{fig:Figure6}
\end{figure*}
In order to use the three qubit register as a quantum memory with error detection, we need at least two entangled qubits. 
Additional single quantum-error correction would require three-partite entanglement \cite{taminiau2014universal, waldherr2014quantum, chang2025hybrid}. 
In general, entanglement between two nuclear spins, i.e. n$_1$ and n$_2$, can be established by using dynamical decoupling to first create a Bell-state between the electron spin and n$_1$ with an additional swap of the electron onto n$_2$ \cite{taminiau2014universal, chang2025hybrid}. 
This approach is limited by the decoherence of the electron spin, since it is in a superposition during register state preparation, as well as infidelities originating from coupling to other nuclear spins due to the electron spin-1/2's unfavorable resonance condition proportional to $(A_\perp^i/\omega_\mathrm{L,n})^2$.\\
Here, in order to circumvent limited coherence of the electron spin, 
we take a different approach entangling two nuclear spins \cite{waldherr2014quantum}. 
We take advantage of the geometric phase $\gamma_\mathrm{g} = \Omega/2$ acquired during a full rotation of the electron spin which is given by half the solid angle $\Omega = 2\pi \, (1-\Delta/\sqrt{\Delta^2 + \Omega_\mathrm{R,e}^2})$ enclosed by the trajectory on the Bloch sphere, where $\Delta$ is the detuning \cite{stas2022robust}.
In the resonant case, i.e. $\gamma_\mathrm{g} \stackrel{\Delta=0}{=}\pi$, we can thus construct a $\pi$-phase-gate conditioned on the nuclear spins' state (CPhase) by driving the respective resonance with a $2\pi$-rotation, see Fig.\ref{fig:Figure6}a.
The circuit diagram of the entangling sequence is depicted in \ref{fig:Figure6}b.
Starting with a SSR on all n$_i$ to prepare the register in $\ket{\Uparrow_{_1}\Uparrow_{_2}\Uparrow_{_3}}$, we then post-select on $\ket{\mathrm{\Uparrow_{_1}\Uparrow_{_2}\Uparrow_{_3}}}$ with an additional SSR window only on that resonance to further improve register initialization to $\FigSixBellFidRegPost$, see \cite{SI}.
After that, we apply unconditional single-qubit $\pi/2$ rotations on n$_1$ and n$_2$ using simultaneous driving of $\omega_{\Uparrow/\Downarrow, \mathrm{n}_i}/2\pi$, 
followed by a CPhase-gate conditioned on $\ket{\mathrm{\Downarrow_{_1}\Downarrow_{_2}\Uparrow_{_3}}}$ with a $\FigFourSwapBWThree$ sinc-pulse. 
The last $\pi/2$-pulse on n$_1$ results in the Bell-state $(\ket{\mathrm{\Downarrow_{_1}}}\ket{\mathrm{\Uparrow_{_2}}} + \ket{\mathrm{\Uparrow_{_1}}}\ket{\mathrm{\Downarrow_{_2}}} )/\sqrt{2}$ between n$_1$ and n$_2$.
It is worth noting that the electron spin has not changed state.
We finally read out n$_1$ and n$_2$ by applying $\FigFourSwapBWTwo$ sinc-pulses at the respective resonance frequencies followed by a laser-pulse, see Fig.\ref{fig:Figure6}b.
Additional single-qubit gates, phase-shifted by $0^\circ$ and $90^\circ$ compared to the initial $\pi/2$-phase, rotate the readout basis to extract the coherences.
From the respective normalized amplitudes $a_i / \sum_i a_i$ we infer the relevant correlators $\langle Z_1 Z_2\rangle$, $\langle Y_1 Y_2\rangle$ and $\langle X_1 X_2\rangle$, see Fig.\ref{fig:Figure6}c and d. 
We then extract a state fidelity to the corresponding Bell-state by calculating $F = (\langle I ,I \rangle + \sum_{i=ZZ, YY, XX} c_i \langle i \rangle)/4$, where $c_i$ depend on the prepared Bell-state. \\
In the given case of $(\ket{\Downarrow_\mathrm{_1}}\ket{\Uparrow_\mathrm{_2}} + \ket{\Uparrow_\mathrm{_1}}\ket{\Downarrow_\mathrm{_2}})/\sqrt{2}$ the correlator coefficients are $c_{ZZ} = -1$, $c_{YY} = +1$ and $c_{XX} = +1$ which results in $F=\FigSixBellFidZeroZero$, demonstrating preparation of an entangled two-qubit state.
Potential error sources to the fidelity are almost entirely dominated by the erroneous $2\pi$ gate fidelity of the $\FigFourSwapBWThree$ sinc-pulse which we set equal to the previously mentioned $\pi$-pulse fidelity $\FigFourSwapFidThree$, see \cite{SI}.
Another relatively minor infidelity arises due to the limited readout fidelity $\FigFourSwapFidTwo$ of the $\FigFourSwapBWTwo$-pulse.
Combined, these two effects set an upper bound for the reachable Bell-state fidelity of $F \approx 0.7$.
\section*{Conclusion and outlook}
In conclusion, we demonstrated bipartite entanglement in a multi-qubit nuclear spin register coupled to a group-IV color center in diamond. 
Due to the quasi-free electron spin of a highly strained SiV we can rely on permanent magnets to supply a stable static magnetic field to sense and control four individual $^{13}$C nuclear spins with $A_\parallel$ down to $\FigThreeAparaFourODMR$. 
In addition to the high sensitivity, continuous decoupling of the electron spin under driving allowed us to implement spectrally narrow sinc-shaped $\mathrm{C}_{\Uparrow_i/\Downarrow_i}\mathrm{NOT}_\mathrm{e}$ and a CPhase-gate for the nuclear-nuclear entanglement as well as single-shot readout on three nuclear spins with high fidelities $F_{\mathrm{init,n}_i} > \FigFiveFinitn$. 
The entanglement can be further improved by using optimal control to increase the fidelity of the CPhase-gate, as has been shown with NV centers \cite{waldherr2014quantum, xie2021beating}. 
This would allow entanglement of all three nuclear spins to implement a logical memory qubit with error correction capabilities. 
Furthermore, the presented SEDOR sensitivity down to coupling strengths of $\FigFourSEDORCouplingTwoOne$ will allow us to increase the nuclear register size further towards a five qubit fault-tolerant logical quantum memory \cite{abobeih2022fault}. 
The thus far limited electron spin initialization fidelity can be improved by increasing the magnetic field strength without drawbacks for the presented direct nuclear spin control and entangling scheme. 
Additionally, enhancing the photon collection by integration of the color centers nano-host into photonic structures will allow single-shot initialization by measurement \cite{antoniadis2023cavity, rosenthal2024single}, as demonstrated here for the nuclear spins.
With the demonstrated long electron spin coherence times, demonstration of spin-photon entanglement comes into reach. 
This can be achieved by integration of the nano-host into optical resonators \cite{antoniuk2024all, lettner2024controlling, bayer2023optical} and using reflection/transmission- \cite{nguyen2019quantum, chan2023chip} or emission-based approaches \cite{ruskuc2025multiplexed, pompili2021realization, fang2024experimental} towards entanglement distribution in a quantum network. Alternatively, the high rates of coherent photons emitted from group-IV color centers combined with efficient photon collection and spin-selective excitation or emission will allow the memory-assisted generation of photonic graph states, enabling long-range quantum communication and measurement-based quantum computation \cite{huet2025deterministic, thomas2022efficient, borregaard2020one, raussendorf2001one}. 

\section*{Acknowledgements}
We thank Matthias Müller for helpful discussion and simulations on using shaped sinc-pulses for enhanced sensitivity. 
We thank Guido van de Stolpe and Philipp Neumann for their fruitful input regarding nuclear spin spectroscopy. 
We thank V.A. Davydov and V. Agafonov for their initial contribution to the nanodiamond material. We thank the Ulm Center for Nanotechnology and Quantum Material for the metal deposition.
The project was funded by the German Federal Ministry of Research, Technology and Space (BMFTR) in the project QR.N (16KISQ006) and by the European Union Program QuantERA in the project SensExtreme (499192368).
\section*{Author information}
\subsection*{Contributions}
AT prepared the sample. 
MK and AT set-up and conducted the experiments.  
MK developed the theoretical models and evaluated the data together with AT. 
DO implemented the FPGA logic.
MK wrote the manuscript together with AT. 
The manuscript was discussed with MK, AT and AK. 
AK supervised the project.
MK and AT contributed equally to this work.

\subsection*{Corresponding author}
Correspondence: alexander.kubanek@uni-ulm.de
\section*{Competing interests}
The authors declare no competing interests.
\newpage
\setcounter{figure}{0}
\onecolumn
\newpage
\printbibliography
	
\end{document}


\maketitle

\subsection*{Supplementary Note 1: Sample and Experimental Setup.}
We are using the same sample as well as the same nanodiamond as in \cite{klotz2025ultra}. 
The nanodiaomonds are grown in a high pressure (8 GPa) and high temperature (1450°C) process, where the silicon is introduced during growth. 
We estimate the size of the nanodiamond to be on the order of 100’s of nm. 
Further details can be found in \cite{antoniuk2024all, klotz2024strongly, lettner2024controlling}.\\
The nanodiamonds are dispersed on a sapphire substrate with a $\SI{200}{\nano \meter}$ thick gold coplanar waveguide (CPW) for microwave supply. 
The nanodiamond is situated in the \SI{10}{\micro\meter} gap of a \SI{50}{\ohm} CPW. 
The sapphire substrate is placed on a custom copper cold finger inside a flow cryostat (Janis ST-500). 
The static magnetic field is applied with four neodymium permanent magnets in Hallbach configuration buried in the cold finger. 
We use a homebuilt 4-f confocal microscope for optical excitation and fluorescence collection, with a high NA objektiv (0.95NA 50x Olympus MPLAPON) at room temperature inside the vacuum chamber of the cryostat. 
Resonant excitation is done with a Ti:Sapphire laser (Sirah), pulsed with an acousto-optical modulator (AOM, G$\&$H 3350-199). 
Resonantly excited photons from the phonon sideband are filtered with a bandpass filter, detected with an avalanche photo diode (Excelitas SPCM-AQRH-14-FC) and time tagged with a TimeTagger Ultra (Swabian Instruments). 
For coherent population trapping (CPT) experiments we modulate the laser with an electro-optical modulator (EOM, JENOPTIK AM705) locked into its transmission maximum (Toptica DLC Pro).\\
All electrical control signals ($\SI{9.415}{\giga \hertz}$ for electron spin (MW), 2.9-4.2$\SI{}{\mega \hertz}$ nuclear spins (RF) and $\SI{350}{\mega \hertz}$ AOM) are synthesized with one analog channel of an arbitrary waveform generator (Keysight M8195A) and split using diplexers (Mini-Circuits ZDSS-3G4G-S+ and ZDPLX-2150-S+). 
To increase the AWG's 8bit dynamic range for low-power electron spin control, we use a digital step attenuator (Analog Devices EVAL-ADRF5700) before amplifying the signal (Mini-Circuits ZVE-3W-183+). 
The RF signal is amplified (Mini-Circuits LZY-22+) and then combined with the MW before the cryostat using another diplexer (Mini-Circuits ZDSS-3G4G-S+). 
For the CPT experiments the microwave is synthesized with the AWG, amplified (Mini-Circuits ZVE-3W-83+) and combined with the lock output using a bias-tee (Mini-Circuits ZX85-12G-S+).\\
To set the attenuation level of the step attenuator we use a digital channel of the AWG to send a number of pulses encoding the attenuation level. 
These are read with an FPGA (AMD Xilinx Zynq 7010, Red Pitaya STEMlab 125-14), which then sets the corresponding bits on the step attenuator. 
The counting of the photons for live feedback is done with the same FPGA. 
The active feedback is realized with a switch (Minicircuits ZASWA-2-50DRA+) controlled by the FPGA, which is deciding after a photon counting window whether a consecutive nuclear spin inverting RF-pulse is passed or dumped.
\newpage

\subsection*{Supplementary Note 2: Additional Electron Spin Characterization.}
\begin{suppfigure}[h]
	\centering
	\includegraphics[scale=1]{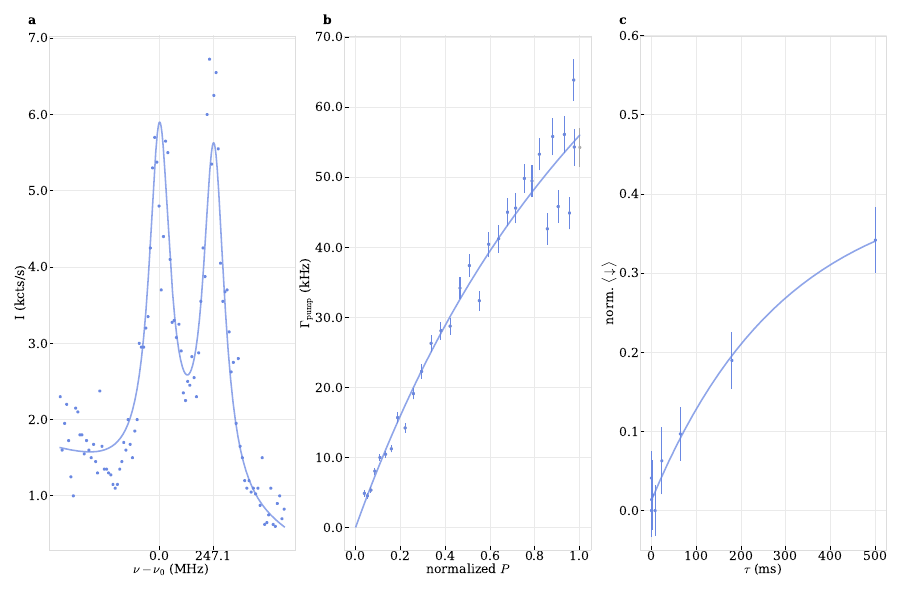}
	\caption{ \textbf{Additional Spin and Optical Properties} 
        \textbf{a} Photoluminescent excitation (PLE) scan where a low-power laser's frequency $\nu$ is scanned across the SiV's spin-split optical dipoles and the respective intensity $I$ in the PSB is collected. 
		\textbf{b} Power-dependent ($P$) spin pumping rate $\Gamma_\mathrm{pump}$ to extract the SiV's optical cyclicity. 
        \textbf{c} Spin relaxation with a pump-probe measurement where pump laser-pulses are interleaved with variable wait times $\tau$.
    }
	\label{figSI:ElectronSpin}
\end{suppfigure}

As described in \cite{klotz2025ultra} we biased the SiV prescreening towards highly strained SiVs by searching for a C-line at wavelengths above typical zero-strain wavelengths (approximately at $\SI{736.9}{\nano\meter}$) and applying a continous-wave (CW) microwave field at the suspected Larmor frequency of a free electron to prevent spin pumping. 
In Supplementary Fig.\ref{figSI:ElectronSpin}a the photo-luminescent excitation (PLE) scan of the SiV is depicted with a resonant CW-MW applied. 
The two spin-conserving transitions, fit with two Lorentzians with a linewidth of \SIFigOneGammaPLE, are split by \SIFigOneSpinSplittingPLE and centered around \SIFigOnePLEcenterwavelength. 
To extract the cyclicity $\eta$ of the SiV, we measure power-dependent spin pumping rates $\Gamma_\mathrm{pump}$ by initializing the SiV electron spin with a resonant laser-pulse on one of the spin-cycling transitions, invert it with a resonant microwave $\pi$-pulse and finally pump it power-dependently. 
We extract $\Gamma_\mathrm{pump}(P)$ as a function of the laser power $P$ from an exponential fit to the fluorescence signal which is then fitted by the expression \cite{klotz2022prolonged, rosenthal2024single}: 
\begin{align}\label{eq:polrate}
\Gamma_\mathrm{P} = \frac{\Gamma_0}{2} \frac{1}{\eta} \frac{s}{1+s}, \qquad s=P/P_\mathrm{sat}.
\end{align}  
where we used a typical excited-state lifetime $1/\Gamma_\mathrm{0} = \SIFigOneOpticalLifetime$ \cite{klotz2025ultra}. 
The procedure resulted in $\eta = \SIFigOneeta$. 
The increased cyclicity compared to a less strained SiV in \cite{klotz2025ultra} is inline with the higher strain and therefore stronger decoupling of the spin and orbital components of the SiV eigenstates.
This also manifests itself in a significantly longer spin lifetime of $T_\mathrm{1} \approx \TimeOne$, see Supplementary Fig.\ref{figSI:ElectronSpin}c, which is an increase of close to three orders of magnitude compared to \cite{klotz2025ultra}.
\newpage
\subsection*{Supplementary Note 3: All-optical nuclear spin sensing.}
\begin{suppfigure}[h]
	\centering
	\includegraphics[scale=1]{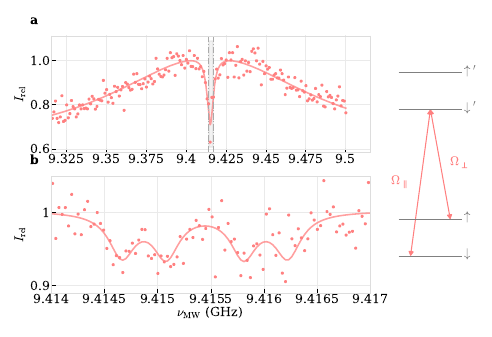}
	\caption{ \textbf{Coherent population trapping} 
		\textbf{a} Resonant driving of the optical $\Lambda$-system consisting of the transitions $\nu_{\uparrow\downarrow'}$ and $\nu_{\downarrow\downarrow'}$. At the Raman condition, when both driving frequencies are resonant, the system is pumped into an optical dark state. Solid line is a double-Lorentzian fit.
        \textbf{b} By reducing the driving strength the hyperfine structure can be sensed optically using CPT, the spectrum is fit by the sum of four Lorentzians (solid line).     
    }
	\label{figSI:CPT}
\end{suppfigure}
We resonantly drive the transition $\nu_{\uparrow\downarrow'}$ with $\Omega_\perp$ and scan another laser frequency, realized by the sideband of a locked electro-optical modulator, with $\Omega_\parallel$ across $\nu_{\downarrow\downarrow'}$ to match the $\Lambda$-system's Raman condition, i.e. $\nu_{\downarrow\downarrow'} - \nu_{\downarrow\uparrow} = \nu_\mathrm{L,e}$, where $\nu_\mathrm{L,e}$ is the electron spin's Larmor frequency.\\
The continuous driving of the $\Lambda$-system results in the preparation of a coherent dark state of the two involved optical dipoles which is decoupled from the driving lasers and hence results in a quench in fluorescence, see Supplementary Fig.\ref{figSI:CPT}a,  called coherent popoulation trapping (CPT).\\
Since the width of the dip is given by residual power-broadening and decoherence in the spin ground-states, we can estimate a decoherence rate $\Gamma_{2,e}^{*, \mathrm{CPT}} = \SIFigTwoGamma$ from the FWHM of a Lorentzian fit to a low-power measurement, Supplementary Fig.\ref{figSI:CPT}b.
We can resolve the strongest coupled nuclear spins all-optically and extract hyperfine coupling components $A_\parallel^1= \SIFigTwoAparaOne$ and $A_\parallel^2= \SIFigTwoAparaTwo$, inline with the measurements presented in the main text.\\
Additionally, the $\Gamma_{2,e}^{*, \mathrm{CPT}}$ is close to the dephasing rate measured by Ramsey interferometry in the main text and the predicted dephasing rate using Monte Carlo simulations for an unpolarized nuclear spin bath. 

\newpage

\subsection*{Supplementary Note 4: Electron spin noise.}
\begin{suppfigure}[h]
	\centering
	\includegraphics[scale=1]{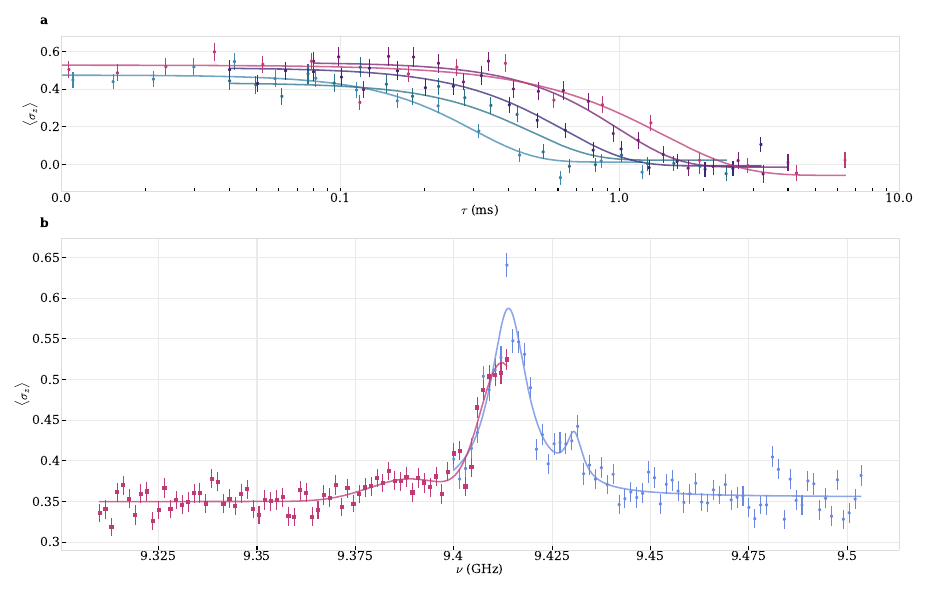}
	\caption{ \textbf{Noise environment.} 
		\textbf{a} CPMG-N type dynamical decoupling with interpulse spacing $\tau$ which is used to fit and extract the respective $T_\mathrm{2,e}^\mathrm{CPMG}$ times from the main text. Solid lines are fits to $a \exp(- (\tau/T_\mathrm{2,e}^\mathrm{CPMG})^\beta) + c$.
        \textbf{b} Double electron-electron resonance measurements. Here $\nu$ is the frequency of the probe $\pi$-pulse, trying to additionally invert target electron spins. Solid lines (red/blue) lines are each fits to the measured data with two Gaussian functions.}
	\label{figSI:DEER}
\end{suppfigure}
The data depicted in Supplementary Fig.\ref{figSI:DEER}a are measurements, where we perform CPMG-type dynamical decoupling. 
For each data trace the number of decoupling pulses $N$ is varied in the sequence $\frac{\pi_\mathrm{x}}{2} - [ \frac{\tau}{2} -  \pi_\mathrm{y} - \frac{\tau}{2}]^N - \left( \frac{\pi_{\pm\mathrm{x}}}{2} \right)$, where $\tau$ is the free evolution time and $\pi_\mathrm{x,y}$ indicate orthogonal rotations of the electron spin on the Bloch sphere.
The last $\pi/2$-pulse is performed around the $\pm$x-axis to read out the electron spin contrast $\sigma_z$.\\
We extract the coherence times $T_\mathrm{2,e}^\mathrm{CPMG}$ from stretched exponential functions with stretching factor $\beta$ in the range of $1.8-4$.
The coherence times scale with $T_\mathrm{2,e}^\mathrm{CPMG} \propto N^\chi $, $\chi = \FigOneChi$ diverting from the expected scaling of $\chi =2/3$ for a nuclear spin bath \cite{medford2012scaling, de2010universal, myers2014probing}.
This behavior can be explained by coupling to additional noise sources such as free electrons \cite{Nguyen2019an, beukers2025control, rosenthal2023microwave}.
To further investigate the noise bath we perform double electron-electron resonance (DEER) spectroscopy by performing a Hahn echo with interpulse spacing $\tau$ on our SiV electron spin and synchronously to the rephasing $\pi$-pulse apply an additional $\pi$-pulse with frequency $\nu$, thereby trying to recouple potential target spins.\\
In Supplementary Fig.\ref{figSI:DEER}b we use a free precession time $\tau = \SI{100}{\micro\second}$ and probe potential spin frequencies below/above (red/blue) the SiV electron spin resonance frequency of $\wLe$.
We measure a broad resonance at $\nu_\mathrm{bath,1} = \FigDEERnuBath$ with FWHM $\Gamma_\mathrm{DEER,1} = \FigDEERFWHMBath$, extracted from a Gaussian fit. 
This agrees well with the expected Larmor frequency of free electrons in the static magnetic field of $\Bext$. \\
Above the resonance of the SiV we measure a second resonance with a loss-of-coherence at  $\nu_\mathrm{DEER,2} = \FigDEERnuSecondSiV$ with a $\Gamma_\mathrm{DEER,2} = \FigDEERFWHMSecondSiV$ which could indicate coupling to a different, nearby defect center.
\newpage

\subsection*{Supplementary Note 5: Numerical system model.}
To evaluate the experimental data we are using a numerical model implemented with QuTIP describing the spin dynamics of our system \cite{qutip1, qutip2}. 
The same model has been used in \cite{klotz2025ultra} and we are only extending the number of nuclear spins we are taking into account. 
We model the system composed of four nuclear spins and one electron spin with the following Hamiltonian ($\hbar = 1$):
\begin{align}\label{SIeqHSys}
\hat{H} = 
&\frac{\omega_\mathrm{L,e}}{2} \hat{\sigma}^\mathrm{e}_{z} +  \sum_{i=1}^{N} \frac{\omega_\mathrm{L,n}}{2}\hat{\sigma}^\mathrm{n}_{z,i} + \hat{\sigma}^\mathrm{e}_{z} \left( \frac{A_\parallel^i}{4} \hat{\sigma}^\mathrm{n}_{z,i} + \frac{A^i_\perp}{4}\hat{\sigma}^\mathrm{n}_{x,i}\right)   \; ,
\end{align}
with the electron/nuclear Larmor frequencies $\omega_\mathrm{L,e/n}$, hyperfine-coupling strengths $A_\parallel^i$ and $A_\perp^i$ and $N$ the number of nuclear spins, see Table \ref{tab:SINM}. 
$\hat{\sigma}_i$ are Pauli operators. 
All operators are understood to be of the right dimension, i.e. with additional identity operators $\mathbb{I}_2$. 
For example $\hat{\sigma}^\mathrm{e}_{z} \equiv  \hat{\sigma}_{z}\otimes\mathbb{I}_2\otimes\mathbb{I}_2\otimes\mathbb{I}_2\otimes\mathbb{I}_2$.\\
Using eq. \eqref{SIeqHSys} we construct a suitable decoherence-free rotation $\exp(-i(\hat{H} + \hat{H}_\mathrm{D})t)$, where for example
\begin{align}
    \hat{H}_\mathrm{D} = \frac{\Omega_\mathrm{R,e}(t)}{2} \left( \cos \phi(t) \hat{\sigma}^\mathrm{e}_{x} + \sin \phi(t)\hat{\sigma}^\mathrm{e}_{y}\right)
\end{align}
is a MW driving term on the electron spin with Rabi frequency $\Omega_\mathrm{R,e}(t)$ and phase $\phi(t)$.\\
In the case of the dynamical decoupling measurements from Fig. 2 of the main text, rectangular-shaped pulses are used, centered in the nuclear spin spectrum, such that $\Omega_\mathrm{R,e}(t) = const. = 2\pi /T_{2\pi}$ and $\phi(t) = \omega_\mathrm{L,e} t + \phi_0$. We then concatenate blocks of the form 
\begin{align} \nonumber
    \hat{U}_\mathrm{DD}(\tau) &= \left[(\hat{U}_{0}(\tau/2) \hat{R}_{x}^N \hat{U}_{0}(\tau/2)\right] \left[(\hat{U}_{0}(\tau/2) \hat{R}_{y}^ {N-1} \hat{U}_{0}(\tau/2)\right]\dots \left[(\hat{U}_{0}(\tau/2) \hat{R}_{y}^2 \hat{U}_{0}(\tau/2)\right]\left[(\hat{U}_{0}(\tau/2) \hat{R}_{x}^1 \hat{U}_{0}(\tau/2)\right]\\ \nonumber
    \hat{U}_{0}(\tau/2) &= \exp(-i\hat{H} \tau/2) \\ \nonumber
    \hat{R}_{x,y} &= \exp(-i(\hat{H} + \hat{H}_\mathrm{D, \phi_0={0,\pi/2}}) T_{2\pi}/2)
\end{align}
to get the evolution $\hat{U}_\mathrm{DD}$ after $N$ blocks as a function of the inter-pulse spacing $\tau$ between two consecutive $\pi$-pulses. Note that $\hat{H}$ is transformed into a frame rotating at $\omega_\mathrm{L,e}$ such that the driving terms become time-independent. \\
We parametrize the initial density matrix by
\begin{align} \nonumber
    \rho_0 = \left(F_\mathrm{e} \ket{\downarrow}\bra{\downarrow} + (1-F_\mathrm{e})\ket{\uparrow}\bra{\uparrow} \right) \otimes \frac{\mathbb{I}_2}{2}\otimes \frac{\mathbb{I}_2}{2}\otimes \frac{\mathbb{I}_2}{2}\otimes \frac{\mathbb{I}_2}{2}
\end{align}
with electron initialization fidelity $F_\mathrm{e}$ and statistical mixture on the nuclear spins.\\
In order to empirically account for loss of coherence we multiply off-diagonal terms of the electron spin's evolved density matrix $\rho_\mathrm{e}(\tau) = \mathrm{tr}_\mathrm{e}\left[\hat{U}_\mathrm{DD}(\tau) \rho_0 \hat{U}_\mathrm{DD}(\tau)^\dagger\right]$ with a function $\exp{(-\tau/\tau_c)^\beta}$ with $\tau_c = \tau_\mathrm{c}^0 N^{\chi-1}$.\\
Specifically, for the 2D XY($\tau$, $N$) measurements from the main text, Fig. 2d, we construct a function with fixed $A_\parallel^i$s calculated from the resonant RF frequencies, experimental parameters $F_\mathrm{e}, T_{2\pi}, \tau, N$ and variable input parameters $\left(\omega_\mathrm{L,n}, \tau_\mathrm{c}^0, \beta, \chi, A_\perp^i\right)$ and fit the dataset to the model with a least-square algorithm. 
In each measurement the number of decoupling pulses $N$ is swept from 20 to 380 in steps of 20.\\
Table \ref{tab:SINM} shows the extracted hyperfine parameters, extracted Larmor frequency $\omega_\mathrm{L, n}$ as well as the necessary experimental parameters, inter-pulse spacings $\tau$ and the electron spin initialization fidelity $F_\mathrm{e}$ extracted from a reference-pulse in each measurement run. \\
These parameters are then used for the simulations of Fig. 2c and f. 
\begin{table}[h]
    \centering
\newpage
    \begin{tabular}{c | c c c c c c}
      $F_\mathrm{e}(\tau)$ & $\tau\,\mathrm{\mu s}$ & & $T_\mathrm{2\pi}$(ns) & $A^i_\parallel / 2\pi$(kHz) & $\omega_\mathrm{L}/ 2\pi$(kHz) & $A^i_\perp/ 2\pi$(kHz) \\
      \cline{1-2} \cline{4-7}
      0.8322 & 8.391  & & 228 & \SINMAparaOne  & \FigTwoomegaLnDD & \FigTwoAperpOneDD   \\
      0.8347 & 8.392 & &  & \SINMAparaTwo  &  & \FigTwoAperpTwoDD   \\
      0.8405 & 8.393 & &  & \SINMAparaThree&  & \FigTwoAperpThreeDD \\
      0.8523 & 8.394 & &  & \SINMAparaFour &  & \FigTwoAperpFourDD  \\
      0.8528 & 8.395 & & &&& \\
      0.8402 & 8.396 & & &&& \\
      0.8341 & 8.397 & & &&& \\
      0.847  & 8.398 & & &&& \\
      0.8302 & 8.399 & & &&& \\
      0.8129 & 8.400 & & &&& \\
      0.8089 & 8.401 & & &&& \\
      0.845  & 8.402 & & &&& \\
      0.875  & 8.403 & & &&& \\
      0.8419 & 8.404 & & &&& \\
    \end{tabular}
    \caption{The parallel hyperfine coupling strengths $A_\parallel^i$, initialization fidelity $F_\mathrm{e}$, Rabi period $T_{2\pi} = 2\pi/\Omega_\mathrm{R,e}$ and inter-pulse spacing $\tau$ are fixed parameters in the fit of the XY($\tau,N$) measurements from the main text Fig. 2d to extract the Larmor frequency $\omega_\mathrm{L,n}$ and the perpendicular hyperfine components $A_\perp^i$.}
    \label{tab:SINM}
\end{table}
\newpage
\newpage
\subsection*{Supplementary Note 6: Low-Power Rabi Measurements.}

\begin{suppfigure}[h]
	\centering
	\includegraphics[scale=1]{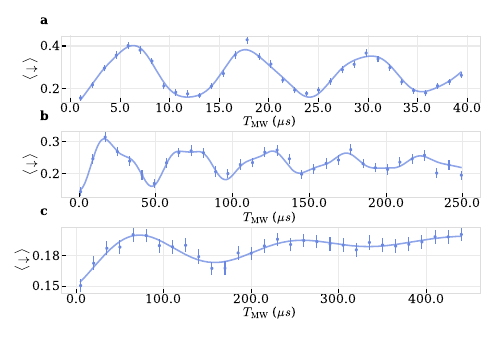}
	\caption{ \textbf{Low-power Rabi measurements} 
		\textbf{a} Ultra-lower power electron spin Rabi oscillations. 
        \textbf{b} Beat of n$_3$
        \textbf{c} Beat of n$_4$
    }
	\label{figSI:LowPowerRabi}
\end{suppfigure}
We use the ability to perform low-power Rabi oscillations on the electron spin to probe the nuclear spin spectrum in the time-domain owing to the long coherence under continuous driving. 
Supplementary Fig.\ref{figSI:LowPowerRabi}a shows a beat in a Rabi measurement, where we first polarize the two strongest nuclear spins and drive Rabi oscillations on the lowest energy electron spin transition. 
From a fit to exponentially damped harmonics, we extract two dominant effective Rabi frequencies of $\Omega_\mathrm{R,e}^\mathrm{eff,1}/2\pi = \SILowPowerRabiFrequencyOne$ and $\Omega_\mathrm{R,e}^\mathrm{eff,2}/2\pi = \SILowPowerRabiFrequencyTwo$, respectively.
We extract an effective detuning $\Delta/2\pi = \sqrt{\Omega_\mathrm{R,e}^\mathrm{eff,2} - \Omega_\mathrm{R,e}^\mathrm{eff,1}} = \SILowPowerRabiDelta $. 
The discrepancy compared to the $A_\mathrm{\parallel}^3 = \FigThreeAparaThreeODMR$, determined in the main text, can be explained by a detuning of both microwave frequencies, for example due to coupling to additional nuclear spins. \\
The same procedure can be repeated by further decreasing the MW-power, see Supplementary Fig.\ref{figSI:LowPowerRabi}b.
Here, we again observe a beating with $\Omega_\mathrm{R,e}^\mathrm{eff,1}/2\pi = \FigThreeBeatRabiOne$ and $\Omega_\mathrm{R,e}^\mathrm{eff,2}/2\pi = \FigThreeBeatRabiTwo$, from which we extract $\Delta/ 2\pi = \FigThreeRabiDetuning$.\\
We are capable of measuring Rabi frequencies down to $\Omega_\mathrm{R,e}^\mathrm{eff}/2\pi = \FigThreeOmegaUltraLowPowerRabi$ with a coherence time of $T_\mathrm{2,e}^\mathrm{Rabi} = \FigThreeTimeTwoUltraLowPowerRabi$, see Supplementary Fig.\ref{figSI:LowPowerRabi}c.

\newpage
\subsection*{Supplementary Note 7: Sinc-pulse fidelity.}
\begin{suppfigure}[h]
	\centering
	\includegraphics[scale=1]{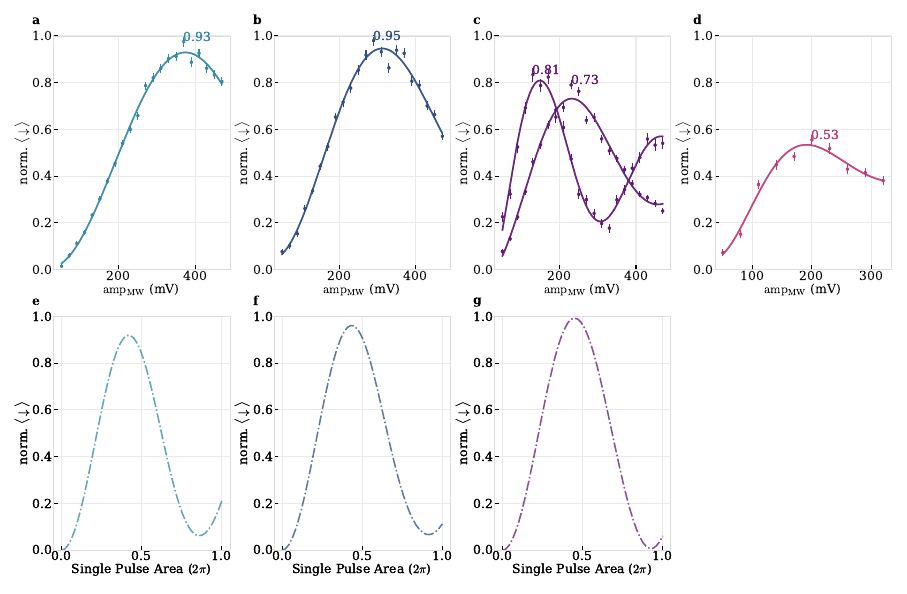}
	\caption{ \textbf{Amplitude sweep of sinc-shaped pulses} 
    In order to find optimal electron spin inversion, we sweep the AWG's MW amplitude $\mathrm{amp}_\mathrm{MW}$ of a truncated sinc-shaped pulse with bandwidths \textbf{a} $\SI{1200}{\kilo\hertz}$, \textbf{b} $\SI{500}{\kilo\hertz}$, \textbf{c} $\SI{150}{\kilo\hertz}$ and \textbf{d} $\SI{40}{\kilo\hertz}$ with additional variable attenuation levels of a digital step attenuator. 
    From an exponentially damped sine (solid line), we extract the respective gate-fidelities.
    For the case of $\SI{150}{\kilo\hertz}$ bandwidth(c), we added a measurement with less MW attenuation to drive a $2\pi$ pulse, used for entanglement generation in the main text. 
    \textbf{e}-\textbf{g} Simulation of the sinc-shaped pulse-amplitude sweep using the numerical model for bandwidths \textbf{e} $\SI{1200}{\kilo\hertz}$, \textbf{f} $\SI{500}{\kilo\hertz}$, \textbf{g} $\SI{150}{\kilo\hertz}$
    We normalize the amplitudes to the whole pulse area of a sinc-pulse.
    }
	\label{figSI:PulseShaping}
\end{suppfigure}
In order to drive nuclear spin conditioned electron spin transitions, we use temporally sinc-shaped pulses ( rectangularly-shaped in the frequency domain) truncated at the second zero-crossing to reduce temporal extent.
The bandwidth $B$ of the pulses are chosen to be approximately $A^i_\parallel/2 \pi$.
In order to realize high and low-bandwidth pulses within one pulse sequence, for example for the strong reference $\pi$-pulse at the beginning of each sequence, we utilize a digital step attenuator.
The latter is controlled in real-time by an attenuation-level encoded pulse from the AWG on the order of 1µs sent to a FPGA which sets the respective level.
Supplementary Fig.\ref{figSI:PulseShaping}a-d show the respective measurement data, where we apply the shaped pulses on all $2^{N_\mathrm{spins}}$ resonances sequentially, where $N_\mathrm{spins}$ is the number of involved nuclear spins, given by the bandwidth of the pulses. 
For example, when $B=\SI{1200}{\kilo\hertz}$, we only use the $2 ^ 1 $ resonances at $(\omega_\mathrm{L,e} \pm A_\parallel^1/2) / 2\pi $, since $B\gg A_\parallel^{2-4}$.
From a fit of an exponentially damped sine we then extract the gate-fidelities.\\
To understand the infidelities, we reproduce the measured data with the numerical model, displayed in Supplementary Fig.\ref{figSI:PulseShaping}e-g.
Since the AWG's output voltage is not calibrated to the experiment, we normalize the amplitudes to the pulse area such that the simulated Rabi frequency becomes $\Omega_\mathrm{R,e}(t)/2\pi = \frac{a}{\int_0^T \mathrm{sinc}_\mathrm{trunc}(t) dt} \mathrm{sinc}_\mathrm{trunc}(t)$, where $\mathrm{sinc}_\mathrm{trunc}(t)$ is the truncated sinc-pulse
and $a$ is the swept amplitude. 
We then add rotating frames during application of the respective $\mathrm{C}_{\mathrm{n}_i}\mathrm{NOT}_e$ frequencies to the system's Hamiltonian \eqref{SIeqHSys} and let the system evolve for a time $T\cdot 2^{N_\mathrm{spins}}$ under driving after which we evaluate the electron spins expectation value and normalize it by $(\langle \downarrow \rangle - (1-F_\mathrm{e}))/(2F_\mathrm{e}-1)$ resulting in Supplementary Fig.\ref{figSI:PulseShaping}e-g.\\
Comparing the simulation for the $\SI{1200}{\kilo\hertz}$ and $\SI{500}{\kilo\hertz}$ pulse with the respective measurements shows good agreement and validates the numerical model.
However, comparison of the $\SI{150}{\kilo\hertz}$ pulse shows a large discrepancy which we attribute to a combination of detuning in the measurement due to additional nuclear spins not taken into account in the system Hamiltonian ($N_\mathrm{spins}=4$) and onset of decoherence during the sinc-pulses $ \SI{26.6}{\micro\second}$.
In addition, one can observe that the maximum population inversion does not happen at a normalized amplitude of $0.5\cdot2\pi = \pi$ but earlier, reflecting a fast over-rotation due to cross-talk.
For the RF nuclear spin driving simulations and experiment in the main text we used the amplitudes of maximum inversion.
\newpage
\subsection*{Supplementary Note 8: Register post-selection.}
\begin{suppfigure}[h]
	\centering
	\includegraphics[scale=1]{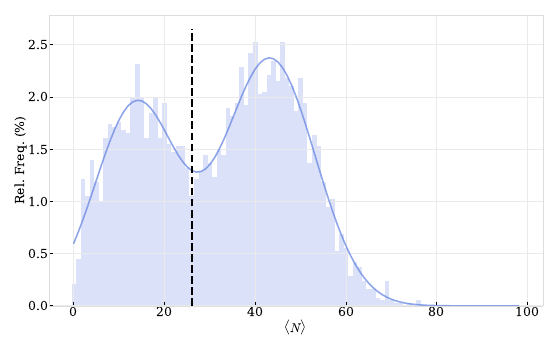}
    \caption{Photon statistics of $\ket{\Uparrow_{_1} \Uparrow_{_2} \Uparrow_{_3}}$, with a double-Gaussian fit to extract a threshold number of photons, black dashed line, which optimizes dark and bright state's mutual overlap. }
	\label{figSI:PostSelection}
\end{suppfigure}
In order to increase the register's initialization fidelity of $\ket{\Uparrow_{_1} \Uparrow_{_2} \Uparrow_{_3}}$ after performing single-shot readout (SSR) on each nuclear spin n$_i$, we additionally apply a SSR with a $\pi$-pulse conditional on $\ket{\Uparrow_{_1}\Uparrow_{_2}\Uparrow_{_3}}$. The SSR parameters are given in the main text.\\
Supplementary Fig.\ref{figSI:PostSelection} shows the respective detected photon number distribution, where we discriminate a dark and bright state with a fidelity of \FigSixBellFidRegPost, which we extracted from a double-Gaussian fit and a subsequent optimized threshold, indicated with a dashed line.

\newpage
\printbibliography